\documentclass{article}
\usepackage{spconf,amsmath,graphicx}
\usepackage{url}
\usepackage{psfrag}
\usepackage{pstool}
\usepackage{textcomp}
\usepackage[absolute]{textpos}

\title{Modeling the HEVC encoding energy using the encoder processing time}
	\name{{ Geetha Ramasubbu \ Andr\'e Kaup \ Christian Herglotz }}   \address{Multimedia Communications and Signal Processing, \\ Friedrich-Alexander University Erlangen-Nürnberg (FAU), Erlangen, Germany\\}

\newcommand{\copyrightstatement}{
    \begin{textblock}{15}(0.3,0.2)    
         \noindent
         \centering
         \textblockcolour{white}
         \footnotesize
         \copyright 2022 IEEE. Personal use of this material is permitted. Permission from IEEE must be obtained for all other uses, in any current or future media, including reprinting/republishing this material for advertising or promotional purposes, creating new collective works, for resale or redistribution to servers or lists, or reuse of any copyrighted component of this work in other works.
    \end{textblock}
}
\begin{document}

\copyrightstatement
%
\maketitle
\begin{abstract}
The global significance of energy consumption of video communication renders research on the energy need of video coding an important task. To do so, usually, a dedicated setup is needed that measures the energy of the encoding and decoding system. However, such measurements are costly and complex. To this end, this paper presents the results of an exhaustive measurement series using the x265 encoder implementation of HEVC and analyzes the relation between encoding time and encoding energy. Finally, we introduce a simple encoding energy estimation model which employs the encoding time of a lightweight encoding process to estimate the encoding energy of complex encoding configurations. The proposed model reaches a mean estimation error of 11.35\% when averaged over all presets. The results from this work are useful when the encoding energy estimate is required to develop new energy-efficient video compression algorithms.

\end{abstract}

\begin{keywords}
video coding, energy-efficiency, HEVC, x265, presets.
\end{keywords}
\vspace{-0.2cm}
\section{Introduction}
\label{sec:intro}
The inception of mobile devices, accessible internet, and various video-on-demand services has increased Internet video traffic. Besides, video-focused social networking services are rising, accounting for further video traffic increase \cite{ciscoWhitepaper1823}. However, this emphasizes enormous storage costs, space needs, and increased server-side energy consumption for video content creation. In addition, the compression methods used for encoding have evolved considerably in recent years. Not only do the modern codecs provide a greater number of compression methods, but their processing complexity has also greatly increased \cite{VVCComplexity}, leading to a significant increase in the energy demand. 

First and foremost, mobile devices are limited in terms of battery power. Notably, video encoding and decoding have considerable power requirements, which poses a problem to battery-powered devices, where the battery drains fast due to increased power requirements. Secondly, the total energy consumption of current coding systems is globally significant as online video contributes to 1\% of total global $CO_2$ emissions. A case study by the Shift Project highlights a considerable energy demand due to the use of various digital equipment in the video processing pipeline and the production of such devices \cite{shiftProject19}. Therefore, we need powerful and energy-efficient video codecs that optimize energy in the encoder-decoder chain and enable energy-aware video-based services in modern video communication applications. 

There exists a large body of literature on decoding energy and optimization, such as a MPEG-1 decoder on a low-power dedicated chip published in 1994 \cite{Molloy94}. Carroll et al. presented an extensive study of the power consumption on smartphones for H.263 \cite{Carroll10} and H.264 encoded bitstreams \cite{Carroll13}. 

An impediment in searching for energy-efficient algorithms is that energy measurements are complex and time-consuming. Hence, we need simple energy estimators to overcome the drawback of complex measurements. There have been several works on modeling the decoding energy in literature, such as in \cite{Li12} and \cite{Raoufi13}. A decoding energy estimation model based on encoded bitstream features, hardware, and software implementation of decoders estimates the true decoding energy with less than 10\% mean estimation error. This model uses various characteristics, such as decoding time in \cite{Herglotz15b}, and characteristics of the sequence such as the number of frames, resolution, and Quantization Parameter (QP) in \cite{Herglotz15a}, and the number of instructions as in \cite{Herglotz17a}. 

Concerning encoding, notably, few recent works have already explicitly addressed the processing energy of the encoder. For example, Rodríguez-Sánchez et al. measured encoding energy \cite{Rodriguez15} and established a relationship between the quantization, as well as spatial details, and coding energy for the intra-only HEVC encoder, but did not take the x265 presets into account. Furthermore, Mercat et al. measure the energy of a software encoder for many different sequences, and encoding configurations \cite{Mercat17}. However, although Mercat et al. consider the various x265 presets, they do not introduce any encoding energy estimation model.

To this end, this work discusses an existing encoding energy estimation model in Section \ref{sec:LR} and introduces an encoding energy measurement setup in Section \ref{sec:setup}. In Section \ref{sec:analysis}, we investigate the encoding process of HEVC-coded bitstreams for various presets and the relation between encoding time and encoding energy and propose a simple encoding energy estimation model to estimate a-priori encoding energies prior to encoder execution in Section \ref{sec:proposal}. Section \ref{sec:eval} presents the results of model evaluation and Section \ref{sec:concl} concludes the paper. 
\section{Literature review of encoding energy estimation model}
\label{sec:LR}
\vspace*{-0.15cm}
This section introduces an existing model for encoding energy estimation. A quantization parameter (QP) based model was introduced for intra-coding in \cite{Rodriguez15}. In this work, Rodríguez-Sánchez et al. introduced a static model to estimate encoding time based on the QP and then used this model to estimate the encoding energy demand. The generalized version of QP based model irrespective of video sequence classes \cite{CTC} as follows: 
\begin{equation}\label{timemodel}t_{\mathrm{enc}}=  \kappa \cdot QP^{3}- \lambda \cdot QP^{2} - \mu \cdot QP+ T_{0},\end{equation}
where $\kappa$, $\lambda$, $\mu$ are the coefficients of the model, QP corresponds to the quantization parameter, and $T_{0}$ is an offset. From this equation, the estimated energy is obtained using the relation with the power and time as:
\begin{equation}\label{QPModel}E_{\mathrm{enc}}=t_{\mathrm{enc}} \cdot P_{\mathrm{avg}},\end{equation}
where $t_{\mathrm{enc}}$ is the estimated time from \eqref{timemodel} and $P_{\mathrm{avg}}$ is the average power consumed during the encoding process. 
This model has several drawbacks. First and foremost, this model was designed for all-intra encoding. Secondly, this static model reports \cite{Rodriguez15} quite accurate estimations for low and high values of QP, i.e., 0–15 and 40–51, but for the mid-range QP values, the errors grow above 15\%. However, the QPs at which the model reports a low error are seldom used. Apart from these, this energy model estimates the energy only for classes B, C, and D of JVET common test conditions \cite{CTC} and 4k sequences were not included. 
\section{Encoding Energy Measurement Setup}
\label{sec:setup}
\vspace{-0.15cm}
For this work, the energy demand of the encoding process is determined with the help of two consecutive measurements, as explained in \cite{Herglotz18c}. Initially, the total energy consumed during the encoding process is measured. And then, the energy consumed in idle mode over the same encoding duration is measured. In the end, the encoding energy $E_{\mathrm {enc}}$ is the difference between these two measurements.
\begin{equation} \label{measure}E_{\mathrm {enc}} =\int _{t=0}^{T} P_{\mathrm {total}}(t)\mathrm{d}t- \int _{t=0}^{T} P_{\mathrm {idle}}(t)\mathrm{d}t,  \end{equation}
where $T$ is the duration of the encoding process, $P_{\mathrm {total}}(t)$ is the total power consumed while encoding, $P_{\mathrm {idle}}(t)$ is the power consumed by the device in idle mode, and t is the time. The confidence interval test for $m$ measurements with a standard deviation of $\sigma$ is defined as follows:
\begin{equation} \label{sta_test}2\cdot \frac {\sigma }{\sqrt {m}}\cdot t_\alpha (m-1) < \beta \cdot E_{\mathrm{enc}}, \end{equation} 
where $\beta$ is the maximum encoding energy deviation, $\alpha$ is the probability with which \eqref{sta_test} is fulfilled and $t_\alpha$ denotes the critical t-value of Students's t-distribution. In this work, we chose $\alpha=0.99$ and $\beta=0.02$ and we repeat the measurements until the condition \eqref{sta_test} is satisfied. Thereby, we ensure that $\bar E_{\mathrm{enc}}$ has a maximum energy deviation of 2\% from the actual energy consumed, i.e., the true encoding energy is not lower than 0.99 times and not higher than 1.01 times the measured mean encoding energy $ \bar E_{\mathrm {enc}}$, with a probability of 99\%. 

In this work, we perform multi-core encoding with the x265 encoder implementation \cite{x265}. We consider 25 sequences from the JVET common test conditions \cite{CTC} with various sequence characteristics such as frame rate, resolution, and content. In addition, we encode the sequences at different x265 presets, which are $\textit{ultrafast}$, $\textit{superfast}$, $\textit{veryfast}$, $\textit{faster}$, $\textit{fast}$, $\textit{medium}$, $\textit{slow}$, $\textit{slower}$, $\textit{veryslow}$, and various Constant Rate Factor (CRF) values, 18, 23, 28, 33. As the encoding energy measurements take a considerable amount of time, especially at slower presets, we consider only the first 100 frames of each sequence. As we use x265's default rate control method CRF, which increases the quantization for high motion frames and lowers it for low motion, therefore, while evaluating the QP based model in Section \ref{sec:LR}, we use average QP, which is obtained at the end of the encoding process.
\section{Encoding Energy Analysis}
\label{sec:analysis}
\begin{figure}[t!]
\centering
 \input{figures/allmeanETEEtilmed.tex} 
 \includegraphics[width=0.5\textwidth]{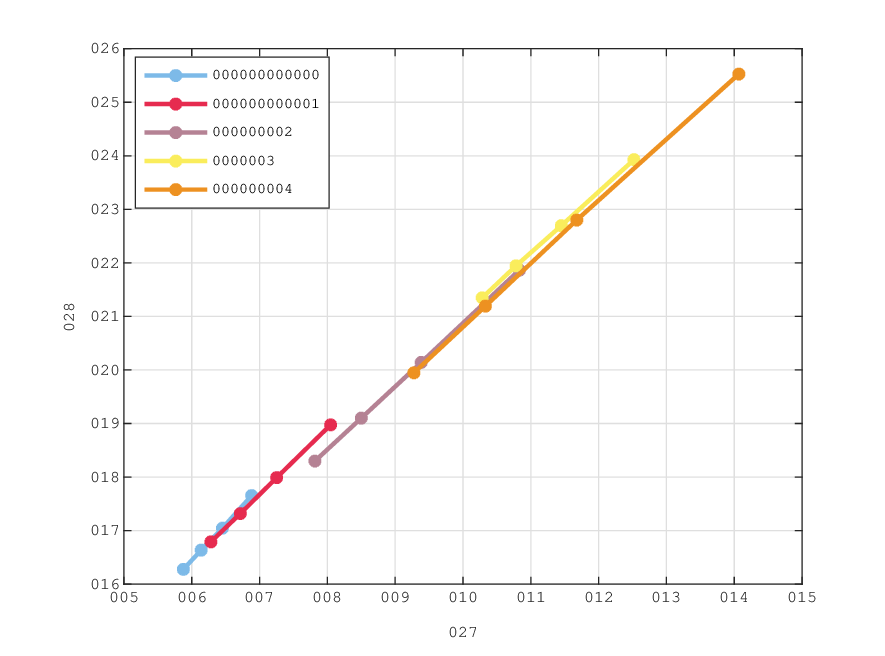}
 \caption{Relation between encoding time and encoding energy for $\textit{ultrafast}$ to $\textit{medium}$ presets, where each marker corresponds to the CRFs 18, 23, 28, 33 from top right to the bottom left.}
 \label{fig:superfast}
 \vspace{-0.4cm}
\end{figure}

\vspace*{-0.15cm}
The energy estimation model from the literature \eqref{timemodel} shows that the encoding time is an important factor in determining the encoding energy. Therefore, we investigate the relation between encoding energy and encoding time. Intuitively, the energy to encode a sequence grows linearly with the encoder processing time, which is experimentally proven by Figure \ref{fig:superfast}. Each colored line in Figure \ref{fig:superfast} represents one of the energy-time relation for presets $\textit{ultrafast}$, $\textit{superfast}$, $\textit{veryfast}$, $\textit{faster}$, $\textit{fast}$, and $\textit{medium}$. In every colored line, each point denotes the average energy-time pair of all considered sequences encoded at CRF values 18, 23, 28, 33. For other x265 presets, we obtain similar characteristics. We report the energy and time values averaged over all the sequences. 

Figure \ref{fig:superfast} shows a linear relationship between processing time and energy, even though multi-core encoding is used. In addition, we can see from Figure \ref{fig:superfast} that, when encoding all the considered sequences at CRF 18, on average, the $\textit{superfast}$ preset requires approximately 0.4 times more energy and processing time and $\textit{fast}$ preset requires approximately 1.7 times as much more energy and processing time compared to that of the $\textit{ultrafast}$ preset at CRF 18. Similarly, $\textit{medium}$ preset requires 2.5 times more energy and processing time when compared to the encoding at $\textit{ultrafast}$ preset at CRF 18. Comparably, $\textit{slow}$ preset requires seven times more energy and 8.5 times more processing time, $\textit{slower}$ preset requires 30 times more energy and 40 times more processing time. These factors decrease at increasing CRF values.

\section{Proposed Encoding Energy Models}
\label{sec:proposal}
\begin{figure}[]
\centering
  \input{figures/allmeanETEEtilmedUF.tex} 
 \includegraphics[width=0.5\textwidth]{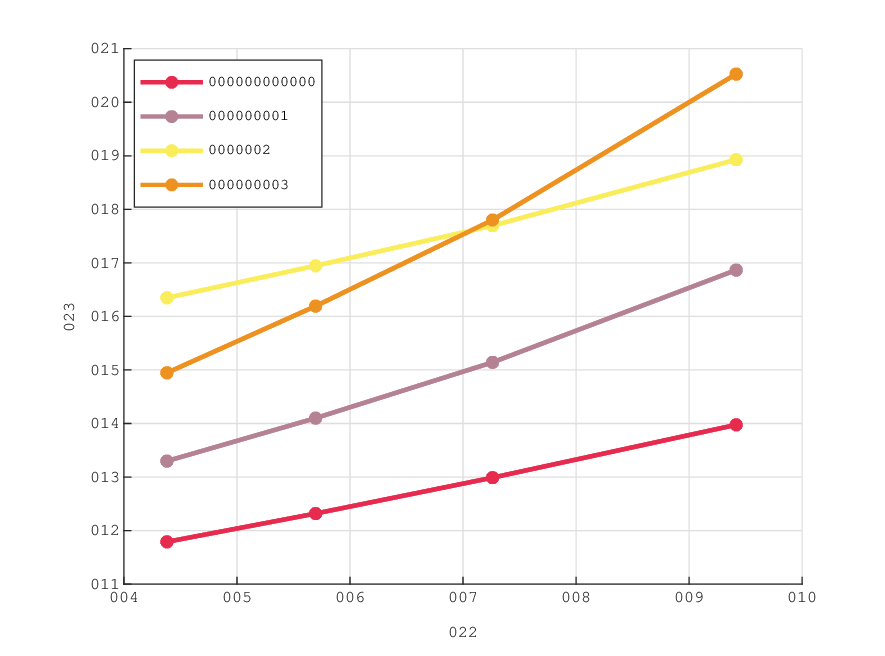}
  \caption{Relation between $\textit{ultrafast}$ encoding time and encoding energy for $\textit{superfast}$ to $\textit{medium}$ presets, where each marker corresponds to the CRFs 18, 23, 28, 33 from top right to the bottom left.}
 \label{fig:allpresets}
\end{figure}

\vspace*{-0.15cm}
Based on the observations presented above, we propose the following model similar to \cite{Herglotz15b} and \cite{Rodriguez15} to estimate the encoder energy demand, which exploits that the encoder processing time  is approximately linear to the encoding energy, such that the energy can be estimated by
\begin{equation}\label{timeModel} \hat E_{\mathrm {enc}} = E_{0} + P \cdot t_{\mathrm {enc}}, \end{equation}
where $t_{\mathrm{enc}}$ is the sequence-dependent encoder processing time. The parameter $P$ (slope) can be interpreted as the linear factor representing the mean encoding power and $E_{0}$ as a constant offset. As a drawback, estimates can only be obtained when the encoding process is executed once on the target device because the encoder processing time needs to be measured. Hence, we adjust this model to allow a-priori energy estimation, i.e., energy estimation without the need to execute the encoder. 
To this end, we investigate the impact of \textit{ultrafast} encoding time on encoding energies of all the presets, as the \textit{ultrafast} encoding time is relatively less costly to obtain when compared to that of other presets. Each colored line in Figure \ref{fig:allpresets} denotes the relation between $\textit{ultrafast}$ encoding time and encoding energy of other presets such as $\textit{superfast}$, $\textit{veryfast}$, $\textit{faster}$, $\textit{fast}$, and $\textit{medium}$. Each point on colored line average energy-$\textit{ultrafast}$ encoding time pair of all considered sequences encoded at CRF value 18, 23, 28, 33.  We report the energy and time values averaged over all the sequences.We can also observe similar linear behavior for presets that are not depicted. Figure \ref{fig:allpresets} shows a linear relationship between $\textit{ultrafast}$ processing time and encoding energy of each preset, with different energy offset and slope.
Based on these observations,  we adapt the model \eqref{timeModel} to estimate the encoding energy for each x265 preset using the encoding time of the $\textit{ultrafast}$ preset i.e., lightweight encoding process, which is less costly to obtain than \eqref{timeModel},  
\begin{equation} \label{model2}\hat{E}_{\mathrm{enc}} = E_{0}+ P \cdot t_{\mathrm{enc,uf}}, \end{equation}
where $E_{0}$ is the offset energy,  $P$ denotes the slope, and $t_{\mathrm{enc,uf}}$ represents the encoding time consumed for a given sequence to be encoded at a given CRF using the ultrafast preset. \vspace{-0.15cm} 
\section{Evaluation}
\label{sec:eval}
\vspace*{-0.15cm}
For evaluation, we choose the mean relative estimation error. 
By doing so, we get more significant results than using the absolute error as we strive to estimate the encoding energy accurately independent of the absolute energy, which can vary by several orders of magnitude. Thus, we show the relative estimation error of the measured encoding energy with respect to the estimated encoding energy for a single bitstream $m$ i.e., each bitstream $m$ corresponds to a single input sequence coded at a specific CRF, and for each preset $X$ as:
\begin{equation} \epsilon_{X,m}=\frac{\hat E_{\mathrm {enc}}- E_{\mathrm {enc}}}{E_{\mathrm {enc}}} \Bigg\vert _{X,m} \end{equation}
where $\hat E_{\mathrm {enc}}$ is the estimated and $ E_{\mathrm{enc}}$ the measured encoding energy from \eqref{model2} and \eqref{measure}, repectively. Then, we calculate the mean estimation error for each preset $X$ over each bitstream $m$ to obtain the overall estimation error for each preset:
\begin{equation} \overline{\epsilon}_{X}=\frac{1}{M}\sum_{m=1}^{M}\vert \epsilon_{X,m}\vert. \end{equation}

In order to determine the model parameter values for each preset, we perform a least-squares fit using a trust-region-reflective algorithm as presented in \cite{Coleman96}. We initially use the measured energies for a subset of the sequences referred to as the training set and their corresponding variables as input, which are the encoding times and the average QPs. As a result,  we obtain the least-squares optimal parameters for the input training set, where we train the parameters such that the mean relative error is minimized. In the end, these model parameters are used to validate the model's accuracy on the remaining validation sequences. The training and validation data set are determined using a ten-fold cross-validation as proposed in \cite{Zaki14}. Using this technique, we randomly divide the complete set of measured energies into ten approximately equal-sized subsets. Then, for each subset, we use the other nine subsets to train the model, and the trained parameters are then used to validate the remaining subset by calculating the relative estimation error for all the sequences.

We tabulate the relative mean estimation error for all the models discussed in this work in Table \ref{tab:mee}. In \cite{Rodriguez15}, the QP based model is trained and validated for each video sequence class separately. Therefore, to have a fair comparison, we evaluate the QP based model for each class and preset. And then, for each preset, we obtain the mean estimation error as the average of per-class mean estimation errors. In addition, even though \cite{Rodriguez15} suggests \eqref{QPModel} for intra encoding, here we use it to train and evaluate inter encoding as our experimental setup is not restricted to intra-only. We consider midrange QP values while encoding, and hence the mean estimation error is large for the QP based model. The average mean estimation error for all the presets is 35.15\% for the QP based model. 

While considering the encoding time-based model, we can see that the encoding time is the better parameter to estimate the encoding energy, and this model has an average mean estimation error of 8.1\%. Notably, this encoding time based model has mean estimation error $<$ 10\% for all the presets except $\textit{slower}$ and $\textit{veryslow}$. For the $\textit{ultrafast}$ encoding time-based model, we can see that the $\textit{ultrafast}$ encoding time can be used as a parameter to estimate the encoding energy with an average mean estimation error of 11.35\%. From these results, it can be interpreted that the encoding time model and the proposed model are relatively precise for all the presets when compared to the QP based model by \cite{Rodriguez15}. 
\vspace{-0.05cm}
Even though the encoding time model has better performance than the proposed model, we cannot use it for a-priori energy estimation as it requires the encoding times for all the presets, whereas the proposed model only requires the $\textit{ultrafast}$ encoding time for estimation. Table \ref{tab:optpara} gives the optimized parameters that are obtained from least-squares fitting. The fitted $P$ values, i.e., slope indicate that $\textit{fast}$ preset consumes 1.2 times more energy than $\textit{superfast}$, medium requires more than twice as much energy as $\textit{superfast}$ and $\textit{veryslow}$ about 40 times as much energy as $\textit{superfast}$.
\begin{table}[]
 \centering
\begin{small}
\caption{Relative mean estimation error for the QP based model \cite{Rodriguez15}, encoding time model, and proposed model for all the bitstreams.}
 \label{tab:mee}
    \begin{tabular}{|l||r |r |r|}
        \hline
         \textbf{Preset} & \textbf{QP based} & \textbf{Encoding time} & \textbf{Proposed}\\
          & \textbf{model \cite{Rodriguez15}} & \textbf{model} & \textbf{model}\\
         \hline \hline
         \textit{ultrafast} &32.82\%  &5.03\%  & -- \\
         \hline
         \textit{superfast} &36.26\%  &4.00\%  & 6.73\%\\
         \hline
         \textit{veryfast} &38.99\%  &8.08\%  &11.39\%\\
         \hline
         \textit{faster} &36.73\%  &6.35\%  &9.56\%\\
         \hline
         \textit{fast} &29.77\%  &4.60\%  &13.21\%\\
         \hline
         \textit{medium} &37.71\%&8.32\% & 10.33\%\\
         \hline
         \textit{slow} &33.53\%  &9.26\%  &11.72\%\\
         \hline
         \textit{slower} &31.81\%  &14.60\%  &12.80\%\\
         \hline
         \textit{veryslow} &38.73\%  &13.39\%  &15.06\%\\
         \hline
         \textbf{Average} &\textbf{35.15\%}  &\textbf{8.18\%}  &\textbf{11.35\%}\\
         \hline
    \end{tabular}
\end{small}
\end{table}
\vspace{-0.25cm}
\begin{table}[]
    \centering
 \centering
\begin{small}
\caption{Parameters $P$, $E_{0}$ fitted for various presets.}
 \label{tab:optpara}
    \begin{tabular}{|c||r |r |}
        \hline
         \textbf{Preset} & \textbf{$P$ (Slope)} & \textbf{$E_{0}$ (Offset)} \\
         \hline
         \textit{superfast} &160.36 &-9.73 \\
         \hline
         \textit{veryfast} &251.67&  -24.28 \\
         \hline
         \textit{faster} &252.48  &-20.15 \\
         \hline
         \textit{fast} &378.68  &-19.02 \\
         \hline
         \textit{medium} &358.21 &-25.51\\
         \hline
         \textit{slow} &1137.65 &-70.20  \\
         \hline
         \textit{slower} &4025.44  &230.18 \\
         \hline
         \textit{veryslow} &6771.39  &425.15 \\
         \hline
    \end{tabular}
\end{small}
\vspace*{-0.5cm}
\end{table}
\vspace*{-0.15cm}
\section{Conclusion}
\label{sec:concl}
\vspace*{-0.15cm}
Energy measurements play a major role in searching for energy-efficient algorithms, but are complex and costly. Therefore, we need valid and simple energy estimators to overcome the drawback of such costly measurements. With this respect, for the HEVC software encoding, it is valid to use the encoding time of $\textit{ultrafast}$ encoding, i.e., encoding time of lightweight encoding process, to estimate the encoding energy. Moreover, the proposed model estimates the encoding energy with a mean estimation error of 11.35\% when averaged over all presets, which outperforms a model from the literature. In addition, obtaining the encoding time for all presets, especially the slower presets, is costly when considering the practical application. Therefore, the proposed model makes more sense than the encoding time model, even with slightly better performance than the proposed model. Furthermore, it should be noted that there is a linear relation between encoding time and encoding energy. Therefore, the minimization of the encoding time is highly correlated with the minimization of the processing energy, which is useful for identifying energy-efficient configurations. We plan to use the sequence properties and encoder configuration to estimate energy and exploit the proposed model for energy-efficient preset decisions in future work.
\vspace{-0.25cm}
\section{ACKNOWLEDGMENT}
\label{sec:aklg}
This work was funded by the Deutsche Forschungsgemeinschaft (DFG, German Research Foundation) – Project-ID 447638564.
\bibliographystyle{IEEEbib}
\bibliography{literature}
\end{document}